%% file: main.tex
\documentclass[12pt]{article}
\usepackage{amsmath}
\usepackage{graphicx}
\usepackage{indentfirst}
\usepackage{subfiles}
\usepackage{enumitem}
\usepackage{lipsum}
\usepackage[numbers]{natbib}
\usepackage{verbatim}
\usepackage{xurl} 

\graphicspath{{images/}{../images/}}

\begin{document}

\subfile{01-titlepage}

\clearpage

\pagenumbering{arabic}

\subfile{02-abstract} 
\subfile{03-purpose} 
\subfile{04-introduction} 
\subfile{05-infrastructure} 
\subfile{06-implementation} 
\subfile{07-evaluation} 
\subfile{08-discussion} 

\bibliographystyle{plainnat}
\bibliography{refs.bib}
\clearpage


\end{document}

%% file: 01-titlepage.tex
\documentclass[../main.tex]{subfiles}

\begin{titlepage}
\begin{center}
\huge \textbf{A Technical Report on Image Classification using AWS}\\
\vspace{96pt}

\large \textbf{Team-4: Members} \\
{Aditya Goverdhana (agoverdh@kent.edu)\\ 
Balakrishna Phani Kommanaboina (bkommana@kent.edu)\\
Naveena Kanderi (nkanderi@kent.edu)\\
Jagadeesh Karri (jkarri1@kent.edu)}

\vspace{12pt}
{\today}\\

\vspace{96pt}
\large{\textbf{KENT STATE UNIVERSITY}}\\
Department of Computer Science\\

\end{center}
\end{titlepage}

\pagenumbering{gobble}

%% file: 02-abstract.tex
\part*{Introduction}
\section*{Problem Statement}
The project aims at building an elastic web application that can automatically scale out and scale in on-demand and cost-effectively by using cloud resources. The resources used were from Amazon Web Services. It is an image classification application exposed as a Rest Service for the clients to access.

\section*{Objectives}
 The application takes the images and returns the predicted output using an image classification model through the AWS resources, an IaaS provider. AWS as an IaaS provider offers a variety of compute, storage and message services. So the tasks involved designing the architecture, implementing RESTful Web Services, a load balancer that scales in and scales out EC2 instances at App Tier according to the demand of the user.
 
\section*{Milestones}
The initial milestones involve
\begin{itemize}
\item Design and build an interactive system for User.
\item Explore and understand the AWS services and improve the infrastructure.
\item Enable AWS Services and run some tests to see storage and computation performance.
\end{itemize}
\newpage
\section*{Tools Used}
\begin{itemize}
    \item AWS services (EC2, SQS, S3).
    \item Pretrained Image classification model.
    \item Web Services (HTML, CSS, Flask).
    \item Testing for resources.
    \item Python for backend of the architecture.
\end{itemize}



%% file: 03-purpose.tex
\part*{Literature Review}
Image Recognition is an essential component of various applications, such as facial recognition, object detection, and product recommendation. With the growing demand for image recognition, many researchers have focused on developing image recognition systems using cloud and edge computing. This literature review summarizes and analyzes five research papers that discuss image recognition as a service.

In the paper "Cloud Strategies for Image Recognition," \citet{9244200} discuss various cloud strategies for image recognition. The authors analyze the features of different cloud providers, such as Amazon Web Services, Google Cloud Platform, and Microsoft Azure, and evaluate their suitability for image recognition tasks. The paper emphasizes the importance of cloud computing in image recognition and provides insights into cloud-based image recognition services.

In "Real-Time Object Detection with TensorFlow Model Using Edge Computing Architecture," \citet{9782169} propose a real-time object detection system using a TensorFlow model and edge computing architecture. The paper presents a system architecture that utilizes edge computing to reduce latency and improve response time. The authors evaluate the system's performance using a dataset of real-world images and demonstrate the effectiveness of the system in detecting objects in real-time.

In "The Utilization of Cloud Computing for Facial Expression Recognition using Amazon Web Services," \citet{9297974} discuss the utilization of cloud computing for facial expression recognition. The authors use Amazon Web Services to develop a facial expression recognition system and evaluate the system's performance using a dataset of facial expressions. The paper demonstrates the effectiveness of cloud-based facial expression recognition systems and highlights the potential of such systems in various applications, such as security and entertainment.

In "Object Detection and Recognition using Amazon Rekognition with Boto3," \citet{9776884} discusses object detection and recognition using Amazon Rekognition with Boto3, a software development kit for Python developers. The paper presents an overview of the Amazon Rekognition service and explains how to use the service with Boto3. The author demonstrates the effectiveness of the Amazon Rekognition service in detecting and recognizing objects in real-time, providing a practical guide for developers.

Finally, \citet{8869300} present "An Efficient Real-time Product Recommendation using Facial Sentiment Analysis." The paper proposes an efficient product recommendation system that utilizes facial sentiment analysis to determine a user's emotional state. The system uses the Microsoft Azure Face API to extract facial features and sentiment scores, which are then used to recommend products. The authors evaluate the system's performance using a dataset of user reviews and demonstrate the system's effectiveness in generating relevant product recommendations.

In conclusion, these five research papers provide insights into image recognition as a service using cloud and edge computing. The papers demonstrate the effectiveness of cloud-based and edge-based image recognition systems and provide practical guidance for developers. The papers also highlight the potential of image recognition in various applications, such as facial recognition, object detection, and product recommendation.

Our project aims to create an elastic web application that uses cloud resources to develop an image recognition application. The project involves designing the architecture and implementing RESTful Web Services to provide access to the application. The use of cloud resources to provide scalable and cost-effective solutions is a common theme in the above 5 research papers. The above research papers emphasize the use of cloud-based services, including AWS services, for scalable and cost-effective image recognition solutions.

\clearpage

%% file: 04-introduction.tex
\part*{Project Background}
Computer vision is a field in artificial intelligence that deals with the analysis and interpretation of visual data, particularly images. One of the primary tasks in computer vision is image classification, where images are categorized into predefined classes or labels through machine learning algorithms. Although similar to image recognition, which also uses machine learning algorithms to interpret visual data, image classification differs in that it categorizes entire images, while image recognition identifies and labels objects within an image.


\begin{figure}[h!]
\centering
\includegraphics[scale=0.4]{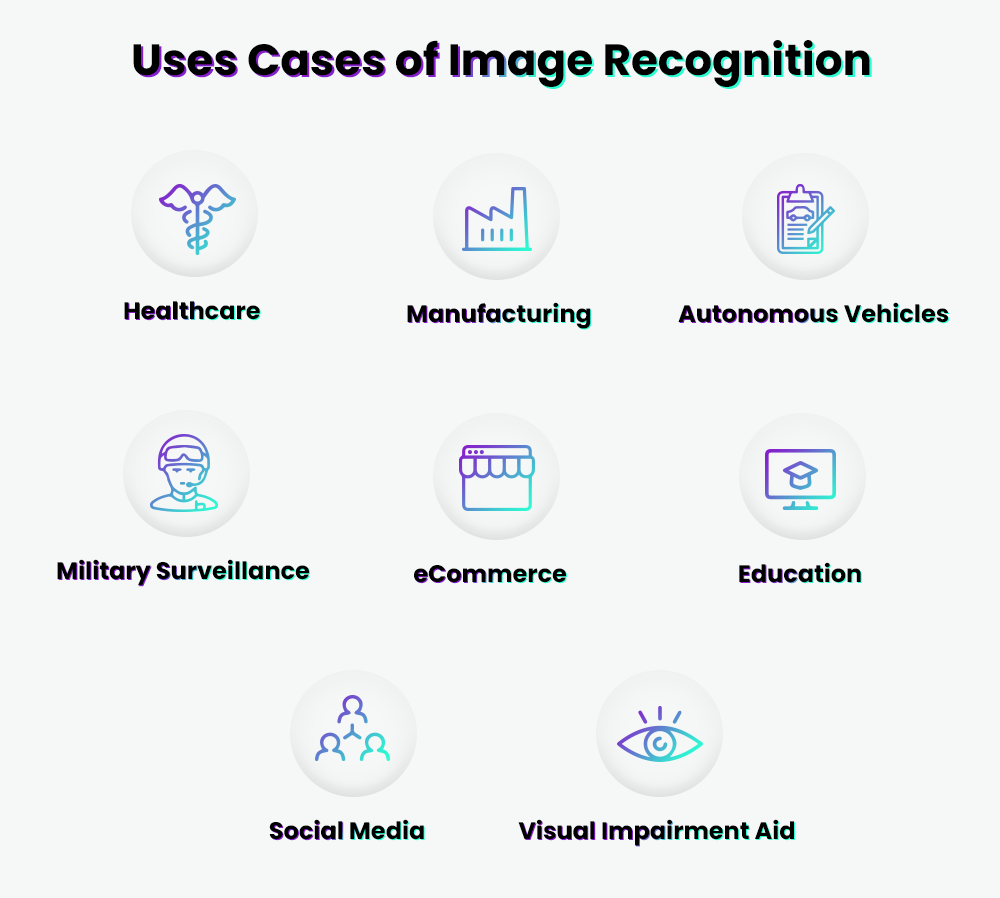}
\caption{Some use-cases of Image Recognition.\citet{im}}
\label{fig:im_uses}
\end{figure}

Image classification, like image recognition, also has a wide range of applications across various industries such as healthcare, automotive, retail, security, and entertainment as shown in the Figure~\ref{fig:im_uses}. Image classification technology can be used for various tasks such as object detection, medical diagnosis, and product recognition. Some examples include:
\begin{itemize}
\item \textbf{Medical Diagnosis:} Image classification can be used for medical imaging analysis, identifying and diagnosing diseases like cancer, or detecting anomalies in medical images.
\item \textbf{Agriculture:} Image classification can be used for crop and soil analysis, plant disease detection, and monitoring crop growth and health.
\item \textbf{Manufacturing:} Image classification can be used for product quality control, identifying defects, and ensuring consistency in production lines.
\item \textbf{Retail:} Image classification can be used for product recognition, inventory management, and customer behavior analysis, such as tracking the popularity of certain products.
\item \textbf{Security:} Image classification can be used for object detection, facial recognition, and surveillance monitoring, such as identifying security threats and detecting suspicious behavior.
\item \textbf{Environmental Monitoring:} Image classification can be used for land use and land cover mapping, identifying deforestation or other changes in the environment, and monitoring wildlife populations.
\end{itemize}

\textbf{IaaS} (Infrastructure as a Service) is a cloud computing model in which cloud service providers offer virtualized computing resources such as servers, storage, and networking to users on a pay-per-use basis. IaaS allows users to scale up or down their computing resources as needed, without having to invest in their own infrastructure. Examples of IaaS providers include Amazon Web Services, Microsoft Azure, and Google Cloud Platform.

IaaS services have a wide range of applications across various industries, such as:
\begin{itemize}
    \item \textbf{Web hosting}: IaaS providers can host websites and web applications in the cloud, providing scalable and reliable infrastructure.
    \item \textbf{Big data}: IaaS providers can provide the computing resources needed to process and analyze large volumes of data.
    \item \textbf{Disaster recovery}: IaaS providers can provide backup and recovery solutions in the event of a disaster or outage.
    \item \textbf{DevOps}: IaaS providers can provide the infrastructure needed for software development and testing.
    \item \textbf{Machine learning}: IaaS providers can provide the computing resources needed to train and run machine learning models.
\end{itemize}

Auto scaling is a key feature of IaaS providers that allows for the automatic adjustment of computing resources based on user demand. Auto scaling ensures that users are not paying for resources that they do not need, and that they have sufficient resources during periods of high demand. This feature can be particularly useful for applications with unpredictable usage patterns, as it enables the system to automatically adjust to changing needs without requiring manual intervention.

Amazon Web Services (AWS) offers several advantages for application monitoring~\citet{rak2011cloud}, load balancing~\citet{9676704}, and Quality of Service (QoS)~\citet{9502607}. Here are some of the key advantages:

Scalability: AWS provides scalable infrastructure that can be easily scaled up or down based on the demand. This is particularly useful for load balancing and QoS as it allows you to dynamically allocate resources to meet the demands of your application.

Availability: AWS provides high availability and fault-tolerant infrastructure that can ensure that your application is always available and responsive. This is important for QoS and load balancing as it helps to minimize downtime and ensure that your application is always accessible.

Monitoring and Analytics: AWS provides a range of monitoring and analytics tools that can help you to monitor the performance and health of your application. This includes tools like CloudWatch, which can help you to monitor your application's performance in real-time and identify any issues or bottlenecks.

Automation: AWS provides a range of automation tools that can help you to automate many of the tasks associated with application monitoring, load balancing, and QoS. This can help you to reduce the workload on your IT team and ensure that your application is always running smoothly.

Cost-effectiveness: AWS is a cost-effective solution for application monitoring, load balancing, and QoS as it allows you to pay only for the resources you use. This can help you to save money on infrastructure costs while ensuring that your application is always running smoothly.

Besides this, AWS Traffic Engineering provides a set of tools and services that allow you to optimize network traffic~\citet{robin2022p4te} and 
improve application performance. With intelligent routing, traffic shaping, load balancing, network performance monitoring, and scalability, AWS TE can help to ensure that traffic is routed to the most optimal endpoint, prevent congestion, evenly distribute traffic, identify and resolve network issues, and accommodate changes in traffic volume and application usage. All of these capabilities can lead to reduced latency, improved resource allocation, and ultimately, better application performance.

Overall, AWS provides a range of advantages for application monitoring, load balancing, and QoS. 
By leveraging the scalable, available, monitoring and analytics, automation, and cost-effective nature of AWS, 
you can ensure that your application is always running smoothly and meeting the needs of your users.

%% file: 05-infrastructure.tex
\part*{Infrastructure}
The infrastructure described is a common architecture for building web applications that require image classification capabilities. The infrastructure is divided into two tiers: the Web-Tier and the Application-Tier.

\begin{figure}[h!]
\centering
\includegraphics[scale=0.68]{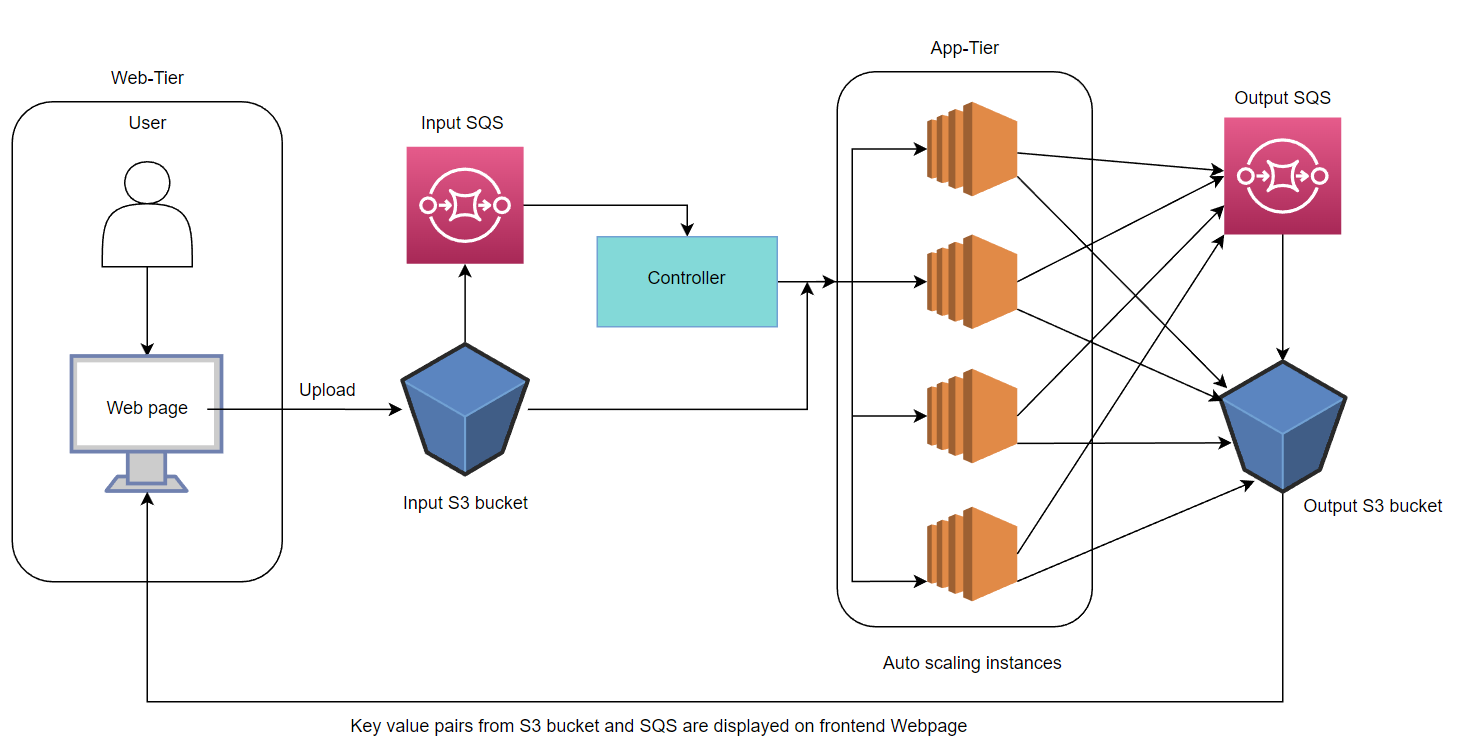}
\caption{System Architecture.}
\label{fig:arc}
\end{figure}

The Web-Tier of the system consists of a simple User Interface, where a user can interact with our system. In our system, user can upload images they want to classify. The Application-Tier contains core functionality of the system, image classification. It also handles business logic, database manipulation functions (CRUD), and communicates with other resources. The overall system architecture of our project can be viewed in the Figure~\ref{fig:arc}.

The number of EC2 instances needed depends on the frequency of data uploads. Scaling in and scaling out of instances are determined based on the number of input images uploaded, which are fed into the SQS queue and the number of messages are calculated. Scaling in is managed by the controller, which automatically terminates instances when there are no more messages left in the input SQS. Scaling out is implemented using the number of running instances and the estimated number of visible SQS messages. Instances are created with the AMI, and the scaling out logic is designed not to create more than 15 instances. To summarize, the number of instances created depends on the number of SQS messages. If there are less than 15 messages, the number of instances created will be the number of messages minus the number of currently running instances. If there are more than 15 messages, 17 instances will run to handle the increased load and avoid errors during the processing of larger images.

This infrastructure utilizes AWS' EC2, SQS, and S3 resources. Specifically, the App Tier relies on S3 and SQS to store and transmit data, respectively. Once the classification results are generated, they are sent as key-value pairs to the output SQS and output S3 bucket to be displayed on the web page.

The infrastructure is a standard two-tier architecture used to develop web applications that incorporate image classification features. The Web-Tier provides a user interface for uploading images, while the Application-Tier contains core functionality, such as image classification, business logic, and database manipulation functions. AWS EC2, SQS, and S3 resources are utilized to create this infrastructure. Scaling in and out of resources is determined by the number of incoming images, which are processed through SQS, and the number of running instances. The infrastructure is designed to create up to 15 instances, and the number of instances created is based on the number of incoming images, with automatic termination of instances when the queue is empty. By leveraging AWS services and this scalable infrastructure, web applications with image classification capabilities can be built in a cost-effective and efficient manner.

%% file: 06-implementation.tex
\part*{Implementation}
The following is a detailed description of the implementation steps involved in building a web application with image classification capabilities on AWS using a two-tier architecture with a Web-tier and App-tier.
\subsection*{Setup}

\subsubsection*{Web-tier:}
To implement the front-end part of the application, we utilized HTML and CSS to design a visually appealing static web page, with HTML providing the structural framework and CSS enabling customization of the layout, font, and color schemes. We integrated the Flask framework to manage communication between the front-end and back-end components, with essential tools and libraries provided to handle various user requests. Using the boto3 libraries, we established a seamless connection between the front-end and App-tier EC2 instance, allowing us to interact with the instance and upload images to the input S3 bucket for processing by the image classification model. Our approach was effective in creating a user-friendly interface while ensuring seamless integration with the App-tier EC2 instance.

\subsubsection*{App-tier EC2 instances:}
App-tier EC2 instances were created following the instructions \citet{ec2} provided by AWS to create an AMI with a t2.small instance type and a root key attached. The default location was set to us-east-1 and the results were stored in JSON format. On the App-tier instances, we installed the necessary libraries listed in "requirements.txt" and added "worker.py" and "image classification.py", which contains a pre-trained ResNet model with ImageNet labels from the Keras library used for image classification. The model has been pre-trained on the ImageNet dataset, which includes over a million labeled images, making it a powerful tool for quickly and accurately classifying images without extensive training on a custom dataset. To run the scripts automatically every minute, we used cron tab and executed the command \texttt{* * * * * python3 /home/ubuntu/worker.py > ./result.txt}.

\subsection*{Execution Steps}

\begin{enumerate}
\item First, we started the Web-tier application manually on the terminal using the command \texttt{python3 app2.py}, which started the Web-tier.
\item Next, the controller is started on the terminal using the command\\ \verb|python3 controller.py|. The script retrieves information about an SQS queue, counts the number of running instances, and starts new instances if necessary. It is designed to be run continuously in a loop as a background process on a local machine.
\item As the local IP address is static and does not change over time, we associated the Web-tier with it. Then, the images are uploaded using the upload button.
\item The required number of instances started running depending on the input number of images.
\item These instances can be on the EC2 dashboard, images were uploaded to the input S3 bucket, and the classification results were uploaded to the output S3 bucket via message passing through SQS.
\item Finally, the results were forwarded to the frontend webpage in key-value pair from SQS and S3 respectively in the format: \verb|image.jpeg, image_label|.
\end{enumerate}

%% file: 07-evaluation.tex
\part*{Results}

\begin{figure}[h!]
\centering
\includegraphics[scale=0.36]{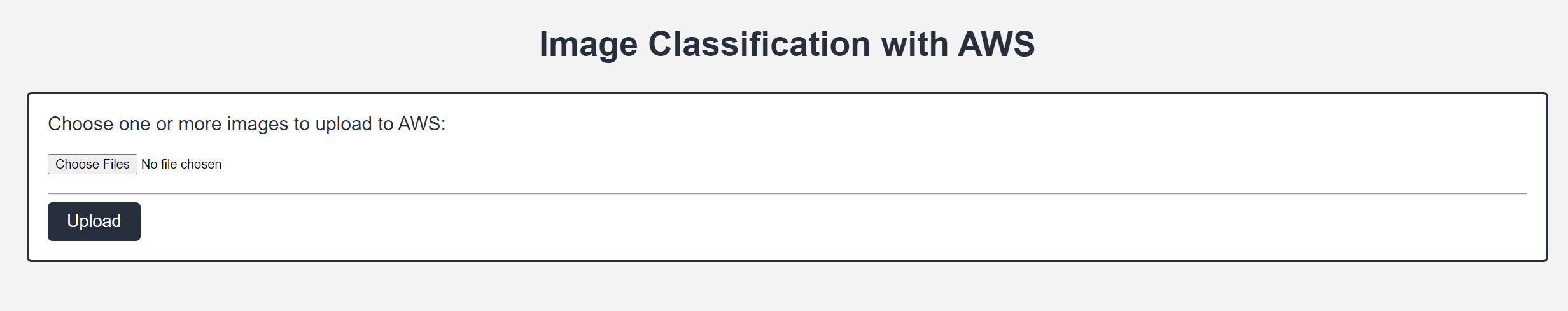}
\caption{User Interface.}
\label{fig:ui}
\end{figure}

The Results section of the document showcases the functionality of the AWS-based image classification system. The user interface of the web application comprises a simple web page that allows users to upload any number of images for classification as shown in Figure~\ref{fig:ui} . The UI has a "Choose file" button that enables users to select the image files to be uploaded, and an "Upload" button to upload them to the AWS services.

\begin{figure}[h!]
\centering
\includegraphics[scale=0.65]{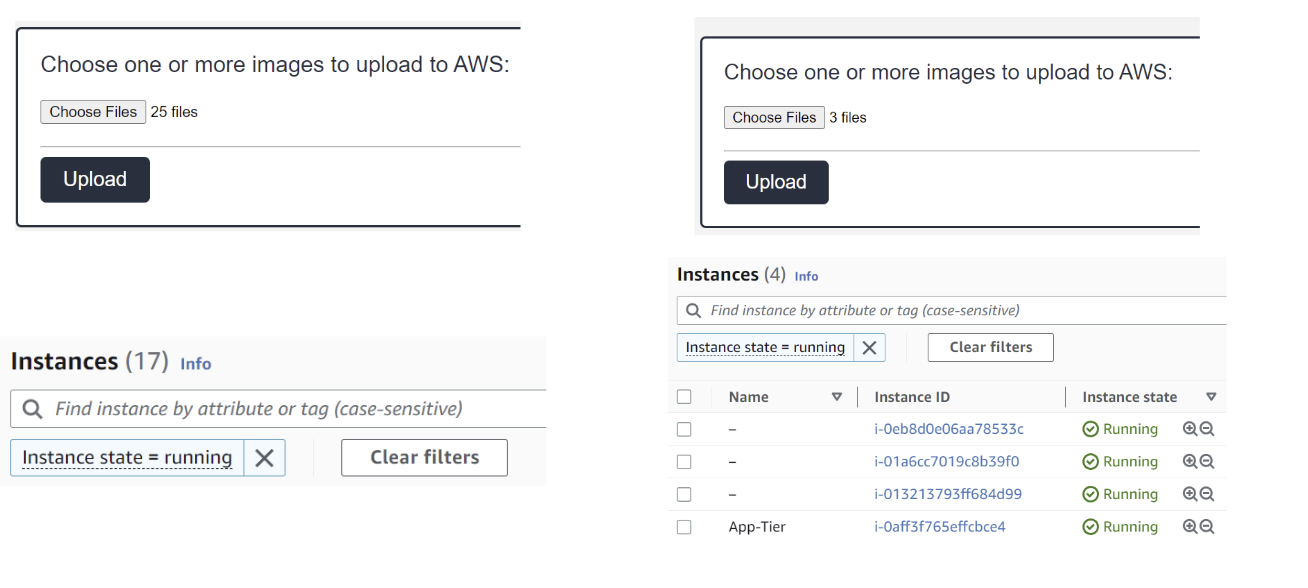}
\caption{If the number of uploaded images is over 15, 17 instances will run, otherwise, the app-tier instance and the number of messages will run for less than 15 uploaded images.}
\label{fig:ins}
\end{figure}

Upon clicking the "Upload" button, the web application automatically creates EC2 instances based on the number of uploaded images. Figure~\ref{fig:ins} displays the number of instances running based on the number of images uploaded. This dynamic scaling of the App-tier instances ensures that the system can handle a large number of requests efficiently.

\begin{figure}[h!]
\centering
\includegraphics[scale=0.36]{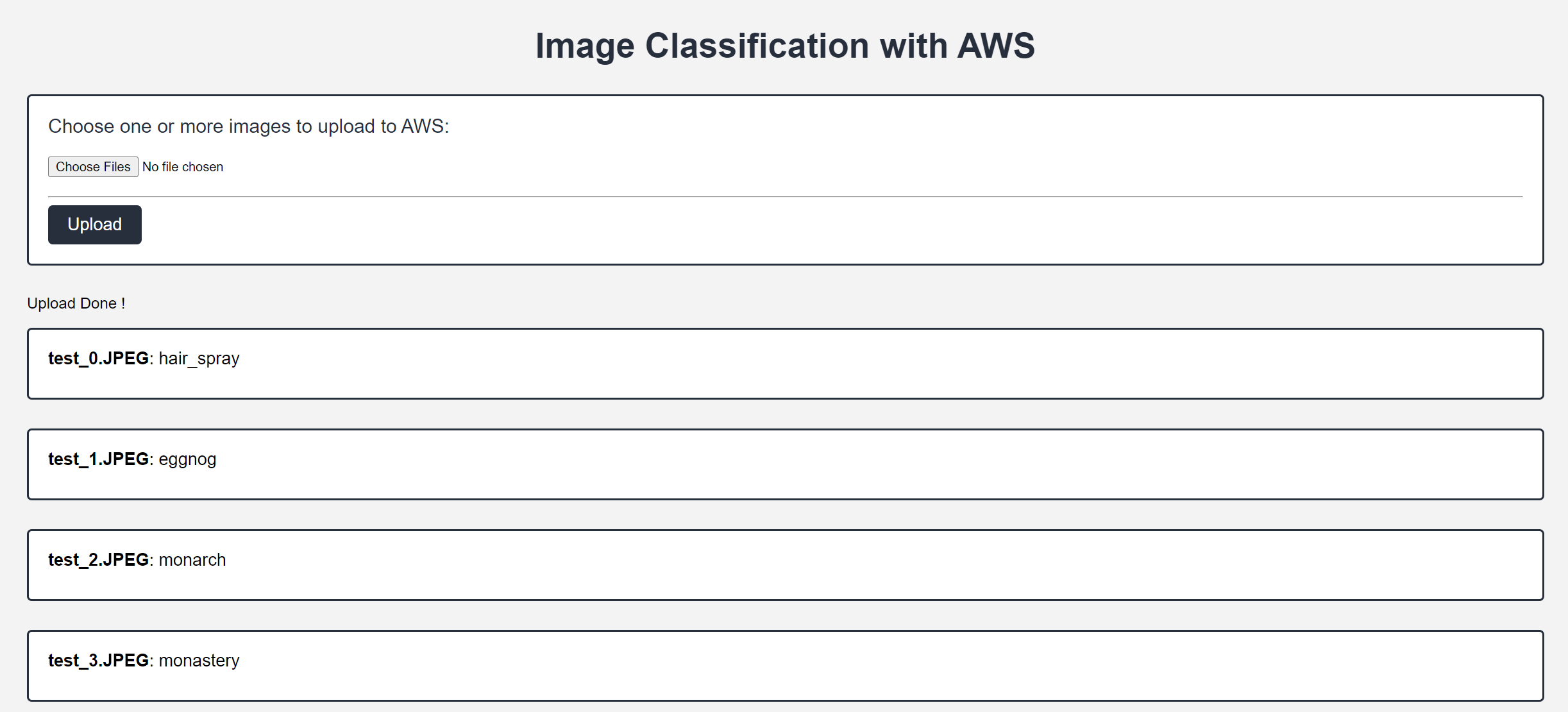}
\caption{Sample output displayed on the UI.}
\label{fig:op}
\end{figure}

After the images have been processed, the system returns the classification results in the form of key-value pairs. The screenshot of Figure~\ref{fig:op} displays the results beneath the "Upload" button on the same webpage. Each image is labeled with its corresponding class and displayed in the format: \verb|test_0.JPEG, hair_spray|. This user-friendly interface enables users to quickly and easily classify multiple images with high accuracy, thanks to the use of a pre-trained ResNet model.

\subsection*{Evaluation based on Metrics:}
In order to evaluate the performance of the system, we measured the response time and boot time for entire processing of the system and booting time of ec2 instance. 
\newpage

\begin{itemize}
    \item Response time for uploading and classifying images was measured using the Python 'time' module.
    \item \textbf{Medium sized images ($\approx 80$ kb):}
    \begin{itemize}
        \item 20 images took an average of 230.15 seconds to display results on the webpage.
        \item 30 images took an average of 215.67 seconds to display results on the webpage.
        \item Medium sized images were classified perfectly.
    \end{itemize}        
    \item \textbf{Lower sized images ($<$2 kb):}
    \begin{itemize}
        \item 27 images took approximately 600 seconds to upload.
        \item Lower sized images did not get accurate classification results.
    \end{itemize}
    \item \textbf{Issues faced while uploading images:}
    \begin{itemize}
        \item Memory errors were encountered while uploading 60 images.
        \item Looping occurred after the 26th image while uploading 40 images.
        \item Investigation needed to determine the maximum number of images that can be uploaded at a time - To determine root cause at varying levels.
    \end{itemize}
    \item Average boot time for App-Tier's EC2 instance (excluding the delay of 30 seconds) was calculated from multiple test runs:
    \begin{itemize}
        \item Average boot time was found to be 71.53 seconds based on the results from the several test runs.
    \end{itemize}
\end{itemize}


%% file: 08-discussion.tex
\part*{Conclusion}

The image classification application implemented in this project using AWS as an IaaS provider was successful in providing users with a platform to upload images for classification. The use of AWS services such as EC2, SQS, and S3, along with Python for the backend, allowed for the scaling of resources based on user demand, providing a cost-effective and scalable infrastructure for image classification applications.

Auto-scaling was implemented as a key feature of the infrastructure, which proved to be useful for applications with unpredictable usage patterns. The project demonstrated the importance of a scalable infrastructure in the field of image classification, which has various applications across industries such as medical diagnosis, agriculture, retail, security, environmental monitoring, and manufacturing.

The evaluation of the image classification system was based on the metrics of response time, boot time and accuracy. The response time for medium-sized images was found to be acceptable, with an average of 215.67 seconds for 30 images. However, the lower-sized images took a longer time to upload, with an average of approximately 600 seconds for 27 images. Additionally, lower-sized images did not get accurate classification results. Some issues were faced while uploading a large number of images, such as memory errors while uploading 60 images and looping after the 26th image while uploading 40 images. Investigation is needed to determine the maximum number of images that can be uploaded at a time. The average boot time for the App-Tier instance (excluding the delay of 30 seconds) was found to be 71.53 seconds based on the results from multiple test runs.

Overall, this project demonstrated the successful implementation of an image classification application using AWS, and provided valuable insights into the architecture and infrastructure required for such applications. The scalability and cost-effectiveness of the infrastructure make it a suitable choice for various image classification applications across different industries.




